%% file: main.tex
\documentclass[fleqn,usenatbib]{mnras}
\usepackage{newtxtext,newtxmath}

\usepackage[utf8]{inputenc}
\usepackage[T1]{fontenc}
\usepackage{ae,aecompl}
\usepackage{graphicx}	
\usepackage{amsmath}	
\usepackage{amssymb}	
\usepackage[normalem]{ulem}
\usepackage{multicol}
\usepackage{pdflscape}
\usepackage{lineno}

\newcommand{\noop}[1]{}

\usepackage{longtable}
\usepackage{array}
\usepackage{rotating}
\newcommand{\PreserveBackslash}[1]{\let\temp=\\#1\let\\=\temp}
\newcolumntype{C}[1]{>{\PreserveBackslash\centering}p{#1}}
\newcolumntype{R}[1]{>{\PreserveBackslash\raggedleft}p{#1}}
\newcolumntype{L}[1]{>{\PreserveBackslash\raggedright}p{#1}}

\input{definitions.tex}

\newcommand{\Nofstars}{48}

\newcommand{\rhk}{R$'_{\rm HK}$}
\newcommand{\bv}{$B-V$}
\newcommand{\radamp}{$\langle A_{\ell=0} \rangle$}

\title[High-mass Kepler RGs II]{The highest mass Kepler red giants--- II. Spectroscopic parameters, the amplitude-activity relation, and unexpected halo orbits}
\author[C. L. Crawford et al.]{
Courtney L. Crawford$^{1}$\thanks{Email: courtney.crawford@sydney.edu.au},
{Yaguang~Li\CNnames{李亚光}}$^{1,2}$,
Daniel Huber$^{1,2}$,
{Jie Yu\CNnames{余杰}}$^{3}$,
\newauthor Timothy R. Bedding$^{1}$,
Sarah L. Martell$^{4}$,
Benjamin T. Montet$^{4}$,
Dennis Stello$^{1,4}$,
\newauthor Howard Isaacson$^{5}$, 
Andrew W. Howard$^{6}$, 
Benjamin J.\ Fulton,$^{6,7}$ 
Jingwen Zhang$^{2,8}$,
\newauthor Alex S. Polanski$^{9,10}$,
Lauren M. Weiss$^{11}$
\\
$^{1}$ Sydney Institute for Astronomy (SIfA), School of Physics, University of Sydney, NSW 2006, Australia\\
$^{2}$ Institute for Astronomy, University of Hawai`i, 2680 Wood-lawn Drive, Honolulu, HI 96822, USA\\
$^{3}$ School of Computing, Australian National University, Acton, ACT 2601, Australia \\
$^{4}$ School of Physics, University of New South Wales, Sydney, NSW 2052, Australia \\
$^{5}$ 501 Campbell Hall, University of California at Berkeley, Berkeley, CA 94720, USA\\
$^{6}$ Cahill Center for Astronomy $\&$ Astrophysics, California Institute of Technology, Pasadena, CA 91125, USA\\
$^{7}$ IPAC-NASA Exoplanet Science Institute, Pasadena, CA 91125, USA\\
$^{8}$ Department of Physics, University of California, Santa Barbara, CA 93106, USA \\
$^{9}$ Lowell Observatory, 1400 W Mars Hill Road, Flagstaff, AZ, 86001, USA \\
$^{10}$ Department of Physics and Astronomy, University of Kansas, Lawrence, KS 66045, USA \\
$^{11}$ Department of Physics and Astronomy, University of Notre Dame, Notre Dame, IN 46556, USA 
}

\date{Accepted xx. Received xx; in original form xx}
\pubyear{2025}
\begin{document}
\label{firstpage}
\pagerange{\pageref{firstpage}--\pageref{lastpage}}
\maketitle

\begin{abstract}

The high-mass (M$>$2 \Msolar{}) Kepler red giant stars are less well-studied than their lower-mass counterparts. In the previous article, we presented a sample of 48 high-mass Kepler red giants and measured their asteroseismic parameters. This article presents spectroscopic measurements from the same sample, using high-resolution Keck/HIRES spectra to determine \Teff{}, [Fe/H], \logg{}, and $v \sin i$. We refined our previous estimates of the stellar masses and radii based on the new \Teff{}. We also examined spectral features that could indicate binary activity, such as the Li line and [C/N] ratios. We found no Li-rich stars or clear [C/N] anomalies, but we observed a correlation between [C/N] and [Fe/H]. We measured chromospheric activity using the $S$-index of the Ca II H \& K lines and found no correlation with internal magnetic fields. However, we confirmed an anti-correlation between surface chromospheric activity and radial mode oscillation amplitudes, which indicates that strong surface magnetic fields weaken stellar oscillations. Finally, we used the Gaia DR3 astrometric data to show that our sample of stars have orbits consistent with all three Galactic kinematic regions. Although these stars are quite young, their orbits carry them into the thick disk and even the halo, raising questions about the accuracy and viability of kinematics in unravelling Galactic history. In future work, we plan to use the spectroscopic parameters measured here to provide better constraints for boutique frequency modelling, which will allow us to test the asteroseismic scaling relations at the high-mass regime.


\end{abstract}


\section{Introduction}

Red giants are important laboratories for many aspects of stellar evolution. When low- and intermediate- mass stars exhaust the hydrogen in their cores, they expand and ascend the red giant branch (RGB), before igniting core He-burning (CHeB), and later ascending the asymptotic giant branch (AGB). The CHeB phase is particularly interesting because its characteristics depend on mass. Stars with masses $<$~2~\Msolar{} will ignite helium explosively on the tip of the RGB (the "helium flash") and move to the "red clump", whereas those with masses $>$~2~\Msolar{} will ignite helium gently and move to the "secondary clump" \citep{Girardi1999_secondaryclump}. Secondary clump stars are much rarer than their lower-mass counterparts due to the Galaxy's initial mass function \citep{Pinsonneault2018_apokasc2, Yu2018_keplercatalog, Hon2021_tess_qlp, Mackereth2021_tess_cvz}. Given the shorter RGB lifetime compared to the CHeB phase, high-mass red giants in that part of the Hertzsprung-Russell diagram are more likely to be in the secondary clump rather than the RGB phase \citep{Crawford2024_highmassrg_paper1}.

Red giant asteroseismology has seen rapid growth since the launch of the Kepler mission, although the majority of work has focused on the lower-mass stars. High-mass red giants are good test beds for several asteroseismic trends that are believed to increase with stellar mass, such as the central frequency and width of the envelope of excited modes 
\citep{Kjeldsen2011_amplitudescaling,Huber2011_solarvalues,Mosser2012_powerexcess,Stello2013_pspacingtracks,Kallinger2014_granulation,Yu2018_keplercatalog} and the relative visibility of the different angular degrees of oscillation \citep{Mosser2012_powerexcess,Fuller2015_suppression,Stello2016_visNature,Stello2016_visPASA,Cantiello2016_suppression}. One particularly interesting aspect of high-mass red giants is their very broad oscillation envelopes \citep{Mosser2012_powerexcess, Yu2018_keplercatalog, Kim2021_width, Sreenivas2024_nuSYD_ARXIVVERSION}. It is currently unclear whether this broadening should be symmetrical around the oscillation frequency of maximum power (\numax{}), which is assumed to scale with the acoustic cutoff frequency in the well-known asteroseismic scaling relations \citep{Brown1991_numaxscaling,Kjeldsen1995_scalingrelations,Belkacem2011_scalingrelations,Hekker2020_scalingrelations_review}. Therefore, these high-mass stars will be very useful in probing the adherence to the \numax{} asteroseismic scaling relation at its extremes. 




This is the second in a series of papers focused on a sample of high-mass ($>$3 \Msolar{}) Kepler red giants. The first paper, \citep[][hereafter Paper~I]{Crawford2024_highmassrg_paper1}, defined the sample of \Nofstars{} stars and described their global asteroseismic parameters, including \numax{}, \Dnu{}, phase shift ($\epsilon$), period spacings ($\Delta\Pi$), and mixed mode visibilities ($V_{\ell=1}$, $V_{\ell=2}$). In this paper, we present a spectroscopic analysis using data from the Keck HIRES instrument. In Section~\ref{sec:obs}, we describe the Keck HIRES setup used to obtain the spectra and describe the determination of spectroscopic parameters (\Teff{}, \logg{}, [M/H], $v\sin i$). In Section~\ref{sec:newmass} we remeasure the masses of our sample using the new spectroscopic parameters, and in Section~\ref{sec:binary_indicators} we explore two potential binary indicators, the Li 6707 \AA{} line and the C/N ratio. In Section~\ref{sec:cahk} we measure the chromospheric activity levels from the Ca H and K lines and show how it relates to the oscillation amplitudes of the stars. In Section~\ref{sec:kinematics} we discuss two stars in our sample (KIC~11518639 and KIC~8378545), which are kinematically consistent with the Galactic halo. Finally, we summarize in Section~\ref{sec:conclusions}. Paper~III will include boutique frequency modelling of some stars with the new constraints we have obtained from their spectra.


\section{Keck HIRES Spectroscopy}
\label{sec:obs}

We collected spectra for all 48 stars in the final sample from Paper~I (listed here in Table~\ref{tab:teff_results_normal}) using the HIRES spectrograph \citep{Vogt1994_hires} mounted on the Keck 10m telescope at Mauna Kea Observatory, Hawai'i. The spectra were obtained and reduced in collaboration with the California Planet Search queue \citep[CPS,][]{Howard2010_cksproject}. The Keck HIRES instrument observes in three wavelength regions: the $b$-band (360--480 nm), the $r$-band (500--650 nm), and the $i$-band (655--800 nm). For determining spectroscopic parameters, we used the $r$-band of the spectra. The spectra have a resolving power of $\sim$52,000 at $600$\,nm and a S/N per pixel of $\sim$100. They were deblazed using a bright early-type (i.e. relatively featureless) spectrum and we stitched together the \'echelle orders using a weighting function that linearly decreases towards the edges of each order.

\begin{figure*}
    \centering
    \includegraphics[width=\textwidth]{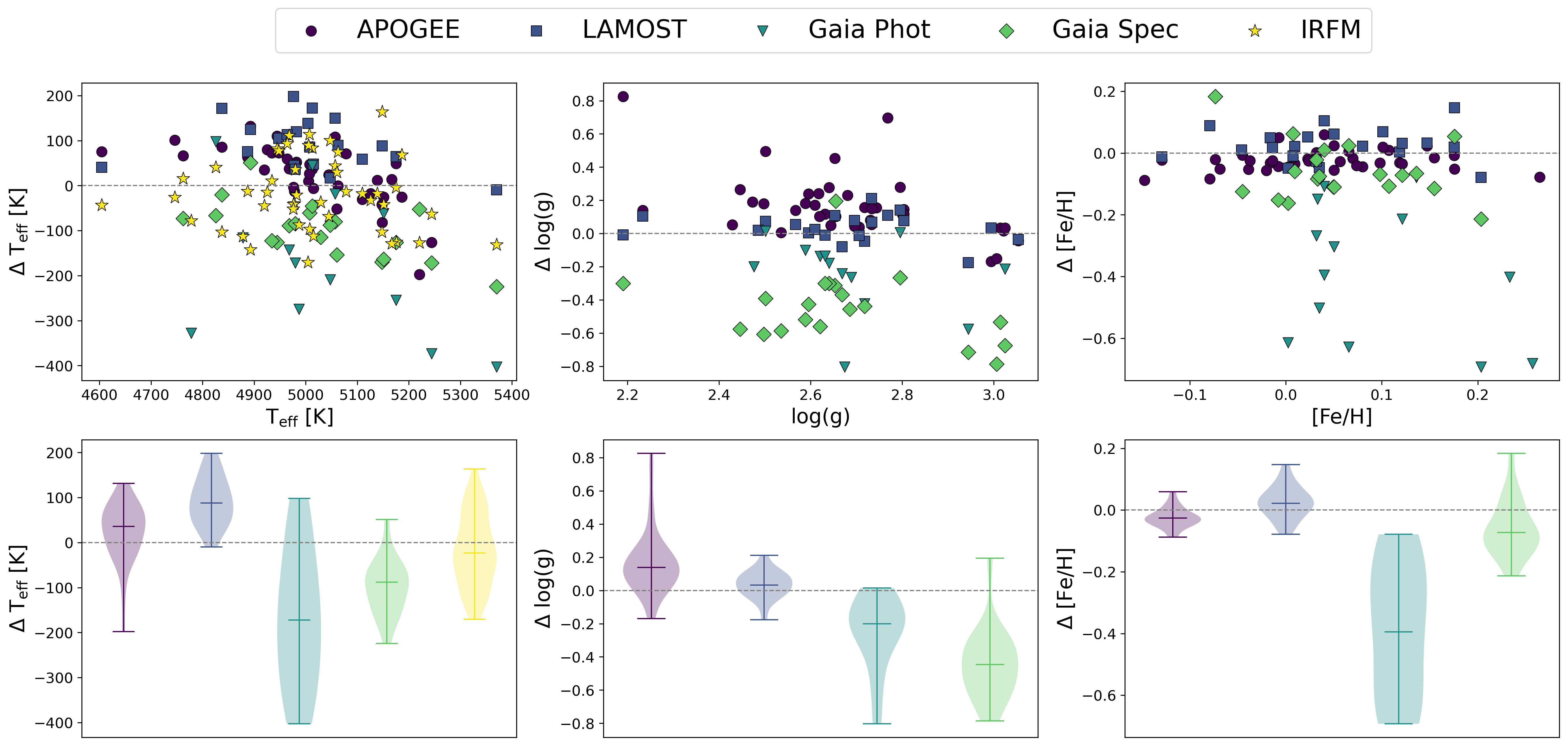}
    \caption{Comparison of our \Teff{} (left column), log(g) (center column), and [Fe/H] (right column) to other surveys and methods: APOGEE in purple circles, LAMOST in dark blue squares, Gaia photometry in light blue triangles, Gaia spectroscopy in green diamonds, and the infrared flux method in yellow stars (\Teff{} only). The upper row shows each star in the sample as a point, and the lower row shows the distribution of the sample differences as a violin plot.}
    \label{fig:spectroscopy_compare}
\end{figure*}

To determine spectroscopic parameters, we used the python tool \texttt{iSpec} \citep{ispec2014,ispec2019}. First, we shifted the spectra to the rest frame by cross-correlating to a template G-type spectrum.  We then used the SPECTRUM radiative transfer code \citep{GrayCorbally1994_SPECTRUM} and MARCS 1D-LTE model atmospheres \citep{Gustafsson2008_MARCSmodels} to generate synthetic spectra, which we then fitted to our data to calculate the spectroscopic parameters \Teff{}, \logg{}, [M/H], and $v\sin i$ using the Fe I and Fe II lines from 530 nm to 635 nm\footnote{Note that non-LTE effects on iron lines are very small in solar-metallicity red giants, and we therefore did not apply a correction to the LTE values \citep{Mashonkina2016_lte_iron,Amarsi2016_lte_iron}.}. We report the uncertainty as the square root of the diagonal elements of the covariance matrix of the fit. The microturbulent and macroturbulent velocities ($\xi_{\rm mic}$ and $\xi_{\rm mac}$, respectively) were automatically derived from the tabulated \Teff{}, \logg{}, and [M/H] using the relations created for the Gaia-ESO survey \citep{GES1,GES2}. 
All spectroscopic parameters are listed in Table~\ref{tab:teff_results_normal}.
In the following subsections and in Figure~\ref{fig:spectroscopy_compare}, we compare our measurements with other sources of spectroscopic parameters to verify consistency. 



\subsection{Comparison with APOGEE and LAMOST}
\label{subsec:apogee}

In Paper~I we adopted spectroscopic parameters from APOGEE \citep{APOGEE,APOGEE_DR16,APOGEE_DR17} and LAMOST \citep{LAMOST}, where available, because those surveys are well-calibrated for a large number of stars. In that analysis, we found that APOGEE and LAMOST showed good agreement (within 2\%), with LAMOST \Teff{} being systematically larger by $\sim$50 K.
In Figure~\ref{fig:spectroscopy_compare} we show a comparison of our spectroscopic parameter estimations for the stars where APOGEE DR17 or LAMOST data is available. For both APOGEE and LAMOST, we see a slight trend in the difference in \Teff{}, where stars with larger \Teff{} have a larger difference. This trend is seen in many spectroscopic surveys, and many of them apply a small \Teff{} correction for this effect (see e.g. \citealt{APOGEE_DR16} Eqn~1). Our \logg{} and [Fe/H] measurements are consistent with both surveys. Overall, our spectroscopic parameters show good agreement with both APOGEE and LAMOST, some of the best estimates available in the literature, which indicates that our spectroscopic analysis is reliable.



\subsection{Comparison with Gaia}
\label{subsec:gaia}

Gaia DR3 reports two different measurements of the spectroscopic parameters--- one based on photometric/colour information, which is reported as \texttt{GSP-phot} \citep{gsp-phot}, and one based on the Gaia RVS (Radial Velocity Spectrometer) spectra, which is reported as \texttt{GSP-spec} \citep{gsp-spec}. In Figure~\ref{fig:spectroscopy_compare} we compare our spectroscopic parameters to those reported by Gaia. The \texttt{GSP-phot} measurements show a much larger scatter than the spectroscopy-based measurements, and are systematically lower on average than all other measurements. The \texttt{GSP-spec} measurements are more consistent with ours, although \texttt{GSP-spec} is again systematically lower. However, \texttt{GSP-spec} reported uncalibrated measurements, and noted a median offset in \logg{} of -0.3 dex from APOGEE DR17 \citep{APOGEE_DR17}, GALAH \citep{Buder25}, and RAVE \citep{RAVE1,RAVE2}. This $-0.3$ dex offset is consistent with the offset shown here, but there are no calibrations recommended for \Teff{}, which is also offset. While our data show poorer agreement with the estimates from Gaia, the differences to photometric measurements are expected, and the differences to the spectroscopic methods are likely due to the much smaller wavelength coverage of the RVS spectra.

Gaia additionally reports rotational velocity information from the RVS spectra in the form of a generalized line-broadening velocity called $v_{\rm broad}$. This value combines all broadening effects, such as $v\sin i$, $\xi_{\rm mic}$, and $\xi_{\rm mac}$. Above roughly 10 km/s, the dominant contribution should be from rotation, and $v_{\rm broad}$ therefore can then be interpreted as a good measure of $v\sin i$ \citep{Fremat2023_vbroad}. As a check, we investigated the consistency between $v_{\rm broad}$ and our iSpec total broadening ($v\sin i + \xi_{\rm mic} + \xi_{\rm mac}$). We found that these velocities are consistent, with a scatter of $\sim$1 km/s. Given that all of our stars have $v_{\rm broad}$ $<$ 13 km/s, this is a roughly 10\% scatter, which is unsurprising given the difficulty of measuring line broadening in giant stars. 

\subsection{Comparison with the IRFM}
\label{subsec:irfm}

Finally, we compared our \Teff{} estimates with those found using the Infrared Flux Method (IRFM) \citep{Blackwell1979_irfm,Casagrande2010_irfmdwarfssubgiants}. We used the {\sc colte} package from \citet{Casagrande2021_irfmcolte}, created for use on GALAH DR4, which uses 2MASS and Gaia bandpass data. The IRFM requires a good estimate of the foreground reddening of each star. For this, we used the python package {\sc mwdust} \citep{Bovy2016_mwdustpackage} to query the \citet{Green2019_3Ddustmap} 3D reddening map using distances reported (via parallax) in Gaia DR3 \citep{GaiaDR3_release}. The {\sc colte} package creates a set of \Teff{} estimates based on multiple different IR colours, and a weighted average of these based on their respective uncertainties. Here, we quote the weighted average \Teff{} calculated by {\sc colte} and show the comparison in Figure~\ref{fig:spectroscopy_compare}. We see no systematic offset from our spectroscopic temperatures, although there is a scatter of up to 200 K, and there is no trend with \Teff{}. The scatter in the relation is due to uncertainties in the reddening values that we adopted.

\begin{table}
	\centering
        \setlength{\tabcolsep}{3pt}
	\caption{Spectroscopic Parameters including \Teff{}, surface gravity, metallicity, projected rotational velocity, and macroturbulent velocity. \label{tab:teff_results_normal}}
	\begin{tabular}{rccccc} 
		KIC & T$_{\rm eff}$ (K) & log(g) & [M/H] & $v\sin i$ (km/s) & $v_{\rm mac}$ (km/s) \\
            \hline
3347458 & 4762 $\pm$ 31 & 2.54 $\pm$ 0.07 & 0.15 $\pm$ 0.03 & 3.44 $\pm$ 0.12 & 3.84 \\
9266192 & 5126 $\pm$ 38 & 3.02 $\pm$ 0.06 & 0.07 $\pm$ 0.03 & 6.01 $\pm$ 0.11 & 3.73 \\
8378545 & 4837 $\pm$ 33 & 2.65 $\pm$ 0.08 & 0.18 $\pm$ 0.03 & 9.83 $\pm$ 0.12 & 3.76 \\
4756133 & 5148 $\pm$ 37 & 3.01 $\pm$ 0.06 & 0.11 $\pm$ 0.03 & 5.07 $\pm$ 0.11 & 3.70 \\
5978324 & 5221 $\pm$ 43 & 2.66 $\pm$ 0.13 & -0.07 $\pm$ 0.03 & 5.92 $\pm$ 0.12 & 4.12 \\
11518639 & 4878 $\pm$ 35 & 2.69 $\pm$ 0.09 & 0.23 $\pm$ 0.03 & 4.98 $\pm$ 0.12 & 3.68 \\
6599955 & 5370 $\pm$ 42 & 2.94 $\pm$ 0.07 & 0.20 $\pm$ 0.03 & 9.24 $\pm$ 0.12 & 3.65 \\
6382830 & 4604 $\pm$ 24 & 2.23 $\pm$ 0.08 & 0.12 $\pm$ 0.02 & 3.15 $\pm$ 0.13 & 4.04 \\
9612933 & 5008 $\pm$ 39 & 2.60 $\pm$ 0.07 & 0.01 $\pm$ 0.03 & 6.83 $\pm$ 0.11 & 4.03 \\
7988900 & 4945 $\pm$ 33 & 2.45 $\pm$ 0.09 & -0.05 $\pm$ 0.03 & 4.50 $\pm$ 0.11 & 4.17 \\
3955502 & 4893 $\pm$ 33 & 2.19 $\pm$ 0.10 & 0.01 $\pm$ 0.04 & 9.24 $\pm$ 0.11 & 4.28 \\
8569885 & 5057 $\pm$ 39 & 2.50 $\pm$ 0.06 & 0.12 $\pm$ 0.03 & 7.70 $\pm$ 0.11 & 3.97 \\
5097690 & 5175 $\pm$ 37 & 2.72 $\pm$ 0.09 & 0.00 $\pm$ 0.03 & 4.67 $\pm$ 0.11 & 3.99 \\
7175316 & 4826 $\pm$ 32 & 2.62 $\pm$ 0.07 & 0.14 $\pm$ 0.03 & 4.31 $\pm$ 0.11 & 3.82 \\
8230626 & 5110 $\pm$ 36 & 3.05 $\pm$ 0.06 & 0.05 $\pm$ 0.03 & 4.61 $\pm$ 0.11 & 3.74 \\
8525150 & 5029 $\pm$ 34 & 2.69 $\pm$ 0.07 & 0.05 $\pm$ 0.03 & 3.39 $\pm$ 0.12 & 3.93 \\
7971558 & 4987 $\pm$ 44 & 2.48 $\pm$ 0.08 & 0.04 $\pm$ 0.04 & 4.31 $\pm$ 0.13 & 4.07 \\
9468199 & 4982 $\pm$ 36 & 2.57 $\pm$ 0.07 & 0.02 $\pm$ 0.03 & 4.36 $\pm$ 0.11 & 4.03 \\
10621713 & 5004 $\pm$ 43 & 2.49 $\pm$ 0.07 & -0.05 $\pm$ 0.04 & 4.13 $\pm$ 0.14 & 4.16 \\
9286851 & 5013 $\pm$ 34 & 2.80 $\pm$ 0.08 & 0.04 $\pm$ 0.03 & 6.16 $\pm$ 0.11 & 3.87 \\
11045134 & 4925 $\pm$ 35 & 2.62 $\pm$ 0.07 & -0.01 $\pm$ 0.03 & 6.09 $\pm$ 0.11 & 4.01 \\
9245283 & 4934 $\pm$ 35 & 2.50 $\pm$ 0.06 & -0.01 $\pm$ 0.03 & 5.38 $\pm$ 0.11 & 4.10 \\
10094550 & 5244 $\pm$ 46 & 2.64 $\pm$ 0.10 & 0.07 $\pm$ 0.04 & 6.98 $\pm$ 0.12 & 3.98 \\
4348593 & 5006 $\pm$ 36 & 2.74 $\pm$ 0.08 & 0.06 $\pm$ 0.03 & 6.25 $\pm$ 0.11 & 3.88 \\
4940439 & 4978 $\pm$ 34 & 2.71 $\pm$ 0.08 & 0.08 $\pm$ 0.03 & 4.39 $\pm$ 0.11 & 3.87 \\
2845610 & 5148 $\pm$ 38 & 2.99 $\pm$ 0.07 & -0.02 $\pm$ 0.03 & 5.24 $\pm$ 0.11 & 3.85 \\
3120567 & 4976 $\pm$ 31 & 2.73 $\pm$ 0.10 & 0.15 $\pm$ 0.02 & 4.39 $\pm$ 0.12 & 3.78 \\
10736390 & 5060 $\pm$ 37 & 3.01 $\pm$ 0.06 & 0.10 $\pm$ 0.03 & 4.15 $\pm$ 0.12 & 3.70 \\
6866251 & 5151 $\pm$ 36 & 3.02 $\pm$ 0.06 & 0.03 $\pm$ 0.03 & 5.02 $\pm$ 0.11 & 3.78 \\
4372082 & 4948 $\pm$ 37 & 2.77 $\pm$ 0.07 & -0.08 $\pm$ 0.03 & 7.90 $\pm$ 0.12 & 4.01 \\
5307930 & 4887 $\pm$ 31 & 2.61 $\pm$ 0.07 & 0.18 $\pm$ 0.03 & 4.36 $\pm$ 0.11 & 3.80 \\
11456735 & 5187 $\pm$ 32 & 2.80 $\pm$ 0.11 & -0.07 $\pm$ 0.02 & 4.46 $\pm$ 0.12 & 4.02 \\
4562675 & 5048 $\pm$ 36 & 2.67 $\pm$ 0.08 & 0.03 $\pm$ 0.03 & 4.04 $\pm$ 0.11 & 3.96 \\
4370592 & 4968 $\pm$ 36 & 2.59 $\pm$ 0.07 & 0.05 $\pm$ 0.03 & 5.38 $\pm$ 0.11 & 3.98 \\
4273491 & 5015 $\pm$ 35 & 2.70 $\pm$ 0.08 & -0.13 $\pm$ 0.03 & 3.09 $\pm$ 0.13 & 4.12 \\
9786910 & 4746 $\pm$ 28 & 2.43 $\pm$ 0.08 & 0.26 $\pm$ 0.02 & 4.09 $\pm$ 0.11 & 3.77 \\
12020628 & 4976 $\pm$ 34 & 3.04 $\pm$ 0.07 & 0.16 $\pm$ 0.03 & 3.82 $\pm$ 0.12 & 3.59 \\
10809272 & 5176 $\pm$ 45 & 2.67 $\pm$ 0.11 & -0.14 $\pm$ 0.04 & 4.20 $\pm$ 0.14 & 4.18 \\
5106376 & 4980 $\pm$ 35 & 2.63 $\pm$ 0.07 & 0.03 $\pm$ 0.03 & 3.51 $\pm$ 0.12 & 3.97 \\
10322513 & 5139 $\pm$ 34 & 2.80 $\pm$ 0.12 & -0.04 $\pm$ 0.04 & 4.46 $\pm$ 0.12 & 3.98 \\
8395466 & 5167 $\pm$ 36 & 2.68 $\pm$ 0.10 & -0.02 $\pm$ 0.04 & 5.17 $\pm$ 0.11 & 4.04 \\
4940935 & 5078 $\pm$ 39 & 2.47 $\pm$ 0.07 & -0.15 $\pm$ 0.03 & 6.90 $\pm$ 0.12 & 4.29 \\
8037930 & 4778 $\pm$ 34 & 2.67 $\pm$ 0.08 & 0.26 $\pm$ 0.03 & 3.41 $\pm$ 0.13 & 3.63 \\
7581399 & 5062 $\pm$ 35 & 2.73 $\pm$ 0.08 & -0.01 $\pm$ 0.03 & 4.27 $\pm$ 0.11 & 3.98 \\
11044315 & 5045 $\pm$ 36 & 2.62 $\pm$ 0.09 & -0.04 $\pm$ 0.03 & 3.58 $\pm$ 0.12 & 4.07 \\
5707338 & 4964 $\pm$ 29 & 2.80 $\pm$ 0.10 & 0.10 $\pm$ 0.02 & 4.01 $\pm$ 0.12 & 3.79 \\
11235672 & 5007 $\pm$ 33 & 2.73 $\pm$ 0.08 & 0.07 $\pm$ 0.03 & 3.00 $\pm$ 0.12 & 3.87 \\
11413158 & 4920 $\pm$ 33 & 2.64 $\pm$ 0.07 & 0.02 $\pm$ 0.03 & 3.91 $\pm$ 0.11 & 3.96 \\
	\end{tabular}
\end{table}

\section{Revised Mass Estimates}
\label{sec:newmass}

\begin{figure}
    \centering
    \includegraphics[width=\columnwidth]{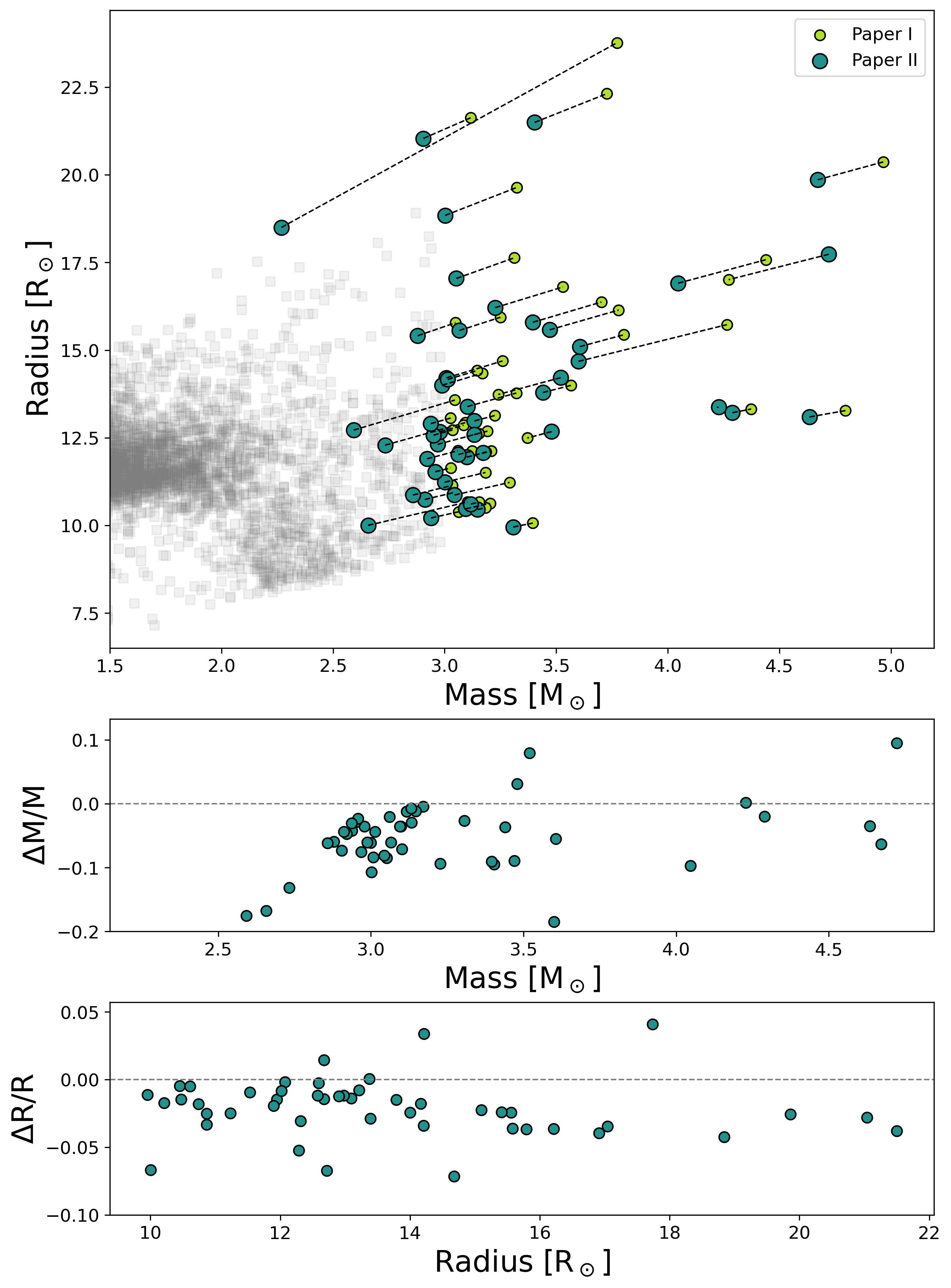}
    \caption{The upper panel shows a radius versus mass plot where in the background (light grey squares) we plot the sample of clump stars from \citet{Yu2018_keplercatalog}. We overplot our sample of stars in coloured circles. Each star has two points, where the smaller green marker shows the values from Paper~I and the larger blue marker shows the mass and radius from this work (having changed the \Teff{}). The change from Paper~I to Paper~II is denoted by black dashed lines for each individual star, where Paper~II refers to this work. The centre panel shows the fractional change in mass from Paper~I to Paper~II versus the mass from this work, with a horizontal dashed grey line denoting no change. The lower panel is the same as the centre panel but using radius instead of mass. The fractional differences are calculated as $\Delta M/M = (M_{\rm Paper~II}- M_{\rm Paper~I})/M_{\rm Paper~II}$ and $\Delta R/R = (R_{\rm Paper~II}- R_{\rm Paper~I})/R_{\rm Paper~II}$. }
    \label{fig:new_keck_masses}
\end{figure}

After homogeneously measuring the spectroscopic \Teff{} for each star, we remeasured the mass for each star using the scaling relation of the form in Eqn~\ref{eqn:mass_scaling}, using the same method as used in Paper~I. 
\begin{equation}\label{eqn:mass_scaling}
\frac{M}{M_{\odot}} = \left(\frac{\nu_{\rm max}}{f_{\nu_{\rm max}}\nu_{\rm max,\odot}}\right)^{3} \left(\frac{\Delta \nu}{f_{\Delta \nu}\Delta \nu_{\odot}}\right)^{-4} \left(\frac{T_{\rm eff}}{T_{\rm eff,\odot}}\right)^{3/2}
\end{equation}
In Figure~\ref{fig:new_keck_masses} we compare the masses and radii calculated in this work to those presented in Paper~I, which used a different \Teff{} but are otherwise the same. It is clear from this diagram that our re-estimation has caused a regression of the values towards the mean of the sample, resulting in smaller masses and radii in most cases. We see a 5\% systematic decrease in mass over all stars in the sample. This caused many of our stars to dip below 3 \Msolar{}, which was the lower mass limit for the sample creation in Paper~I, but since it is not lower than the threshold for secondary clump stars, we have not removed the stars from the sample here. We also see an average $\sim$2\% decrease of the radii of all stars.
All masses and radii are reported in Table~\ref{tab:massrad_actamp}.


As described in Paper I, we used {\sc asfgrid} \citep{Sharma2016_fdnu,Stello2022_asfgridextension} to calculate the correction factor for \Dnu{} ($f_{\Delta \nu}$), used to calculate both mass and radius, with the assumption that our stars are core He-burning (CHeB). However, as demonstrated by \citet{Hon2024_fdnu_issues}, the $f_{\Delta \nu}$ values generated by {\sc asfgrid} seem to produce discrepant results at its extremes. Using the $f_{\Delta \nu}$ values from {\sc asfgrid} in the scaling relations results in combinations of mass and radius which lie outside the range of model values (see their Figure~11).
\citet{Hon2024_fdnu_issues} also specifically discussed two of the highest mass stars in our sample, KIC~3347458 and KIC~4756133, and showed using different evolutionary tracks from {\sc asfgrid} itself that the \numax{} and \Teff{} of these stars seem more consistent with significantly lower masses (see their Figure~15).
In Paper~III, we will include further discussion on proper estimation of $f_{\Delta \nu}$ based on independent mass estimates and how this value influences the scaling relations.

\begin{table}
    \centering
    \setlength{\tabcolsep}{5pt}
    \caption{Mass, Radius, Activity, and Amplitudes \label{tab:massrad_actamp}}
    \begin{tabular}{rcccc}
         KIC & Mass (M$_\odot$) & Radius (R$_\odot$) & $S$-index &  $\langle A_{\ell=0} \rangle$ (ppm)\\
         \hline
3347458 & 4.67 $\pm$ 0.33 & 19.86 $\pm$ 0.49 & 0.123 & 22.98 \\
9266192 & 4.63 $\pm$ 0.21 & 13.09 $\pm$ 0.21 & 0.265 & 8.40 \\
8378545 & 4.05 $\pm$ 0.52 & 16.91 $\pm$ 0.73 & 0.327 & 16.53 \\
4756133 & 4.29 $\pm$ 0.39 & 13.21 $\pm$ 0.40 & 0.233 & 10.42 \\
5978324 & 4.72 $\pm$ 0.30 & 17.73 $\pm$ 0.39 & 0.233 & 14.59 \\
11518639 & 3.60 $\pm$ 0.49 & 14.68 $\pm$ 0.67 & 0.222 & 15.04 \\
6599955 & 4.23 $\pm$ 0.20 & 13.38 $\pm$ 0.23 & 0.124 & 11.16 \\
6382830 & 2.27 $\pm$ 0.46 & 18.50 $\pm$ 1.26 & 0.128 & 48.80 \\
9612933 & 3.61 $\pm$ 0.93 & 15.10 $\pm$ 1.30 & 0.177 & 14.84 \\
7988900 & 3.47 $\pm$ 0.30 & 15.58 $\pm$ 0.48 & 0.125 & 19.42 \\
3955502 & 3.40 $\pm$ 0.23 & 21.50 $\pm$ 0.64 & 0.117 & 36.20 \\
8569885 & 3.40 $\pm$ 0.16 & 15.79 $\pm$ 0.30 & 0.134 & 20.98 \\
5097690 & 3.44 $\pm$ 0.19 & 13.79 $\pm$ 0.30 & 0.122 & 16.55 \\
7175316 & 3.23 $\pm$ 0.32 & 16.21 $\pm$ 0.56 & 0.193 & 18.89 \\
8230626 & 3.31 $\pm$ 0.14 & 9.95 $\pm$ 0.15 & 0.223 & 7.45 \\
8525150 & 3.48 $\pm$ 0.20 & 12.68 $\pm$ 0.25 & 0.132 & 13.89 \\
7971558 & 3.00 $\pm$ 0.27 & 18.84 $\pm$ 0.65 & 0.092 & 32.10 \\
9468199 & 3.10 $\pm$ 0.18 & 13.39 $\pm$ 0.28 & 0.117 & 13.97 \\
10621713 & 3.05 $\pm$ 0.36 & 17.04 $\pm$ 0.71 & 0.098 & 27.41 \\
9286851 & 3.04 $\pm$ 0.19 & 10.87 $\pm$ 0.24 & 0.262 & 8.63 \\
11045134 & 3.01 $\pm$ 0.44 & 14.21 $\pm$ 0.70 & 0.213 & 15.19 \\
9245283 & 3.07 $\pm$ 0.21 & 15.56 $\pm$ 0.39 & 0.116 & 23.31 \\
10094550 & 3.52 $\pm$ 0.23 & 14.21 $\pm$ 0.32 & 0.173 & 12.90 \\
4348593 & 3.13 $\pm$ 0.22 & 12.98 $\pm$ 0.32 & 0.283 & 13.33 \\
4940439 & 3.10 $\pm$ 0.28 & 11.95 $\pm$ 0.37 & 0.144 & 14.60 \\
2845610 & 3.09 $\pm$ 0.26 & 10.47 $\pm$ 0.30 & 0.265 & 8.39 \\
3120567 & 2.97 $\pm$ 0.30 & 12.31 $\pm$ 0.43 & 0.158 & 13.25 \\
10736390 & 3.17 $\pm$ 0.23 & 12.07 $\pm$ 0.31 & 0.246 & 10.39 \\
6866251 & 3.15 $\pm$ 0.15 & 10.45 $\pm$ 0.20 & 0.230 & 9.30 \\
4372082 & 3.00 $\pm$ 0.12 & 11.23 $\pm$ 0.19 & 0.141 & 11.17 \\
5307930 & 2.99 $\pm$ 0.18 & 14.00 $\pm$ 0.30 & 0.092 & 19.97 \\
11456735 & 3.12 $\pm$ 0.24 & 10.61 $\pm$ 0.28 & 0.163 & 10.37 \\
4562675 & 3.13 $\pm$ 0.18 & 12.59 $\pm$ 0.25 & 0.124 & 15.75 \\
4370592 & 3.01 $\pm$ 0.32 & 14.16 $\pm$ 0.51 & 0.184 & 19.68 \\
4273491 & 3.06 $\pm$ 0.15 & 12.02 $\pm$ 0.20 & 0.141 & 14.20 \\
9786910 & 2.90 $\pm$ 0.21 & 21.04 $\pm$ 0.69 & 0.121 & 38.61 \\
12020628 & 2.66 $\pm$ 0.18 & 10.00 $\pm$ 0.23 & 0.286 & 10.14 \\
10809272 & 2.73 $\pm$ 0.17 & 12.29 $\pm$ 0.27 & 0.121 & 18.69 \\
5106376 & 2.98 $\pm$ 0.39 & 12.67 $\pm$ 0.57 & 0.119 & 15.84 \\
10322513 & 2.94 $\pm$ 0.11 & 10.21 $\pm$ 0.15 & 0.126 & 10.86 \\
8395466 & 2.92 $\pm$ 0.23 & 11.90 $\pm$ 0.32 & 0.118 & 13.39 \\
4940935 & 2.88 $\pm$ 0.20 & 15.41 $\pm$ 0.41 & 0.143 & 20.13 \\
8037930 & 2.59 $\pm$ 0.18 & 12.72 $\pm$ 0.31 & 0.077 & 24.86 \\
7581399 & 2.91 $\pm$ 0.23 & 10.74 $\pm$ 0.29 & 0.132 & 11.15 \\
11044315 & 2.95 $\pm$ 0.19 & 12.57 $\pm$ 0.32 & 0.129 & 16.83 \\
5707338 & 2.86 $\pm$ 0.14 & 10.87 $\pm$ 0.20 & 0.219 & 10.08 \\
11235672 & 2.96 $\pm$ 0.13 & 11.53 $\pm$ 0.20 & 0.125 & 15.03 \\
11413158 & 2.94 $\pm$ 0.15 & 12.90 $\pm$ 0.24 & 0.115 & 18.67 \\
    \end{tabular}
\end{table}

\section{Potential Binary Indicators}
\label{sec:binary_indicators}

Stellar multiplicity is known to increase with stellar mass \citep{Lee2020_binaryfraction, Offner2023_stellarmultiplicityreview}. According to \citet{Moe2017_multiplicity}, 81 $\pm$ 6 \% of stars with masses from 3 to 5 \Msolar{} on the main sequence have at least one companion, while 36 $\pm$ 8 \% have at least two companions. Binarity, especially close binarity, is known to weaken the oscillations of red giant stars, often suppressing them completely \citep{Gaulme2014_binarysuppression,Schonhut-Stasik2020_binarysuppression}. It is natural to expect that the likelihood of binary interactions, including mass transfer, tidal spin-up, and mergers, should increase with the multiplicity. However secondary clump stars do not reach as large of radii as lower mass stars on the red giant branch of evolution, and thus are less likely to interact with nearby companion stars. Regardless, understanding which stars in our sample have possibly had binary interactions (especially those which may have gained or lost mass) is paramount to a performing a precise test of the asteroseismic scaling relations, which we will be doing in Paper~III. 

It is not always straightforward to identify whether or not a star is in a binary system. In Paper~I, we discussed various possible indications of binarity from the time-series data, including Gaia parameters, binary signals in the light curve, and possible contamination of the light curves from nearby stars. We did not find any clear indications of binaries or contamination from that work. In the following sections, we discuss two potential indicators of binary interactions using the spectroscopic data: Li-enrichment and [C/N] ratios. 



\subsection{Lithium}
\label{subsec:lithium}

Li-enrichment in red giants is an unsolved puzzle in stellar evolution, even in the era of large spectroscopic surveys. LAMOST spectra have shown that approximately 1\% of all red giants are Li-rich, defined as having an abundance A(Li)~$\geq$~1.5 \citep{Casey2019_lirichgiants,Gao2019_lirichgiants}. Nearly 80\% of Li-rich red giants are in the red clump, where they are undergoing core He-burning \citep{Casey2019_lirichgiants,Singh2021_superlirich}. Secondary clump stars exhibit Li-enrichment more often than primary red clump stars, with occurrence rates of 3.6\% and 1.6\%, respectively \citep{Martell2021_lisrc_rate}. Based on a larger occurrence of anomalously fast rotators and a simultaneous enhancement of $s$-processed material found using GALAH spectra \citep{Sayeed2024_lirichgiants}, one of the most promising formation mechanisms for the Li-rich giants is tidal spin-up due to an intermediate-mass AGB companion. Therefore, we investigated the rate of Li-richness in our sample as a potential indicator of binary interactions on the upper RGB.

\begin{figure*}
    \centering
    \includegraphics[width=\textwidth]{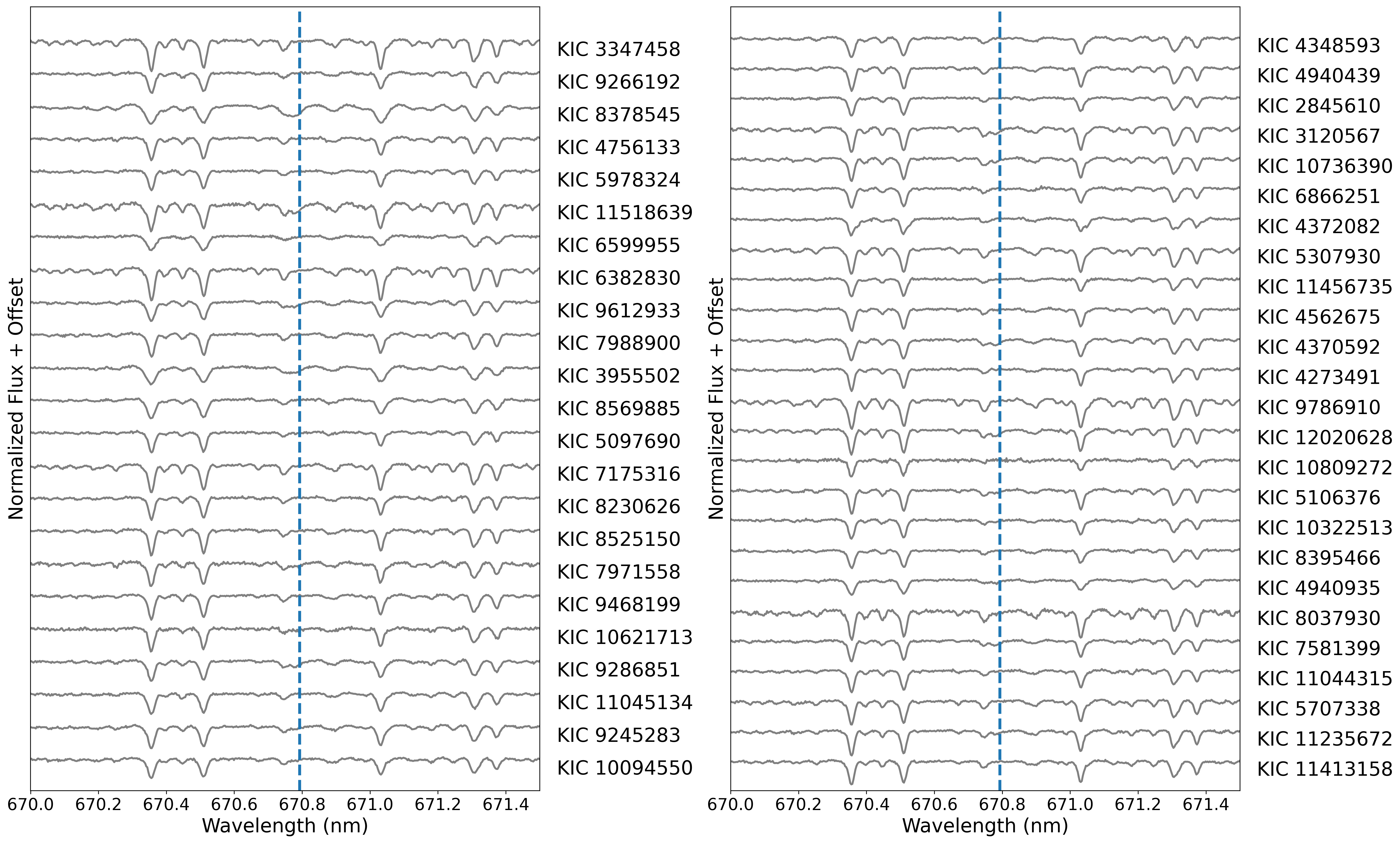}
    \caption{Here we show the Li 670.7926 nm line region of each of our stellar spectra. The vertical blue dashed line denotes the location of the Li line center. All spectra are in the rest frame and have been continuum-normalized.}
    \label{fig:lithium}
\end{figure*}

We show the Li 6707.926 \AA{} region of the $i$-band spectra for all of our targets in Figure~\ref{fig:lithium}. There is no evidence of Li-enhancement in any of our high-mass red giant stars, and none of the Li lines are strong enough to measure a lithium abundance. This is not unusual, however. Using the 3.6\% Li-richness occurrence rate for secondary clump stars from \citet{Martell2021_lisrc_rate}, we would expect to find only one or two stars with enhanced surface lithium in our sample of 48 stars. 
Ultimately, our sample is too small and the enrichment mechanism is not understood well enough for further interpretation of the lack of Li-rich stars in our sample.




\subsection{C/N Ratio}
\label{subsec:cn_ratio}

The surface [C/N] ratio of red giants reflects the depth of the first dredge-up at the base of the RGB, which depends strongly on stellar mass. As a result, [C/N] is commonly used as an indication of the stellar mass for red giants which, in conjunction with the stellar evolutionary stage, can then be used to estimate the age \citep{Martig2016_CNratios}.
Many outliers from the [C/N]-mass relation for CHeB stars show signs of binary interactions \citep{Bufanda2023_CNoutliers}. However, a star's observed [C/N] ratio also depends on its birth [C/N], which can be traced through its [Fe/H]. \citet{Roberts2024_CNvsFeH} demonstrated that much of the variation in [C/N] for giant stars is linked to birth [C/N] and [Fe/H], with more pronounced differences at the higher mass end of the [C/N]-mass relation. They provided a functional form for the [C/N]-mass-[Fe/H] relation in CHeB stars, which extends to higher masses than RGB stars (their Eqn.~12 and Table~3). However, the empirical [C/N]-mass relation is only well studied up to masses of $\sim$2.5 \Msolar{} due to a lack of data for high-mass red giant stars. Here, we investigate the [C/N] ratios of our sample derived by APOGEE \citep{APOGEE_DR17} versus stellar mass for signs of potential binary interactions (as in \citealt{Bufanda2023_CNoutliers}) but additionally note that the [C/N]-mass relation for high-mass red giants is not well constrained.

\begin{figure}
    \centering
    \includegraphics[width=\columnwidth]{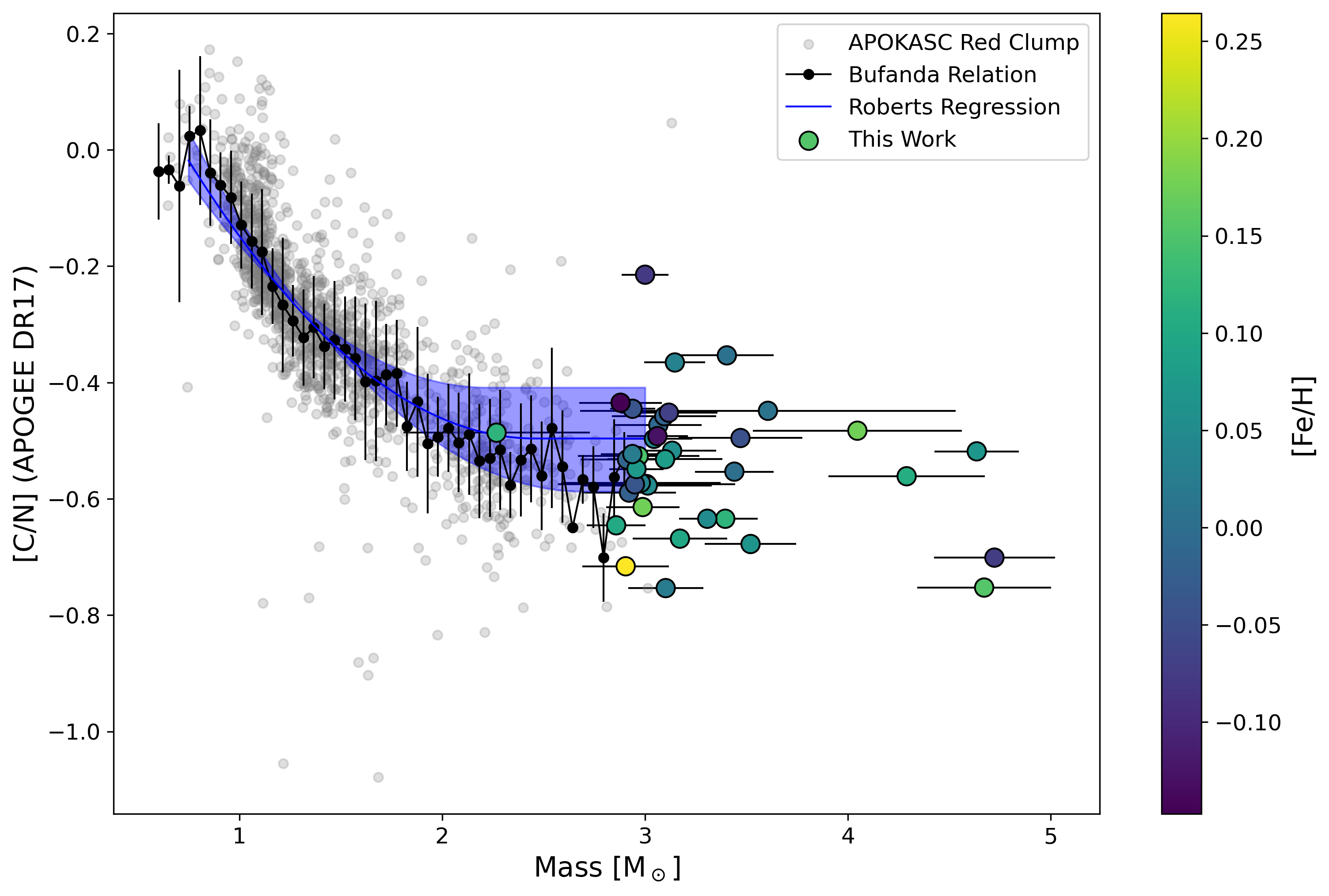}
    \caption{In the background (light grey circles), we plot CHeB stars from the APOKASC-2 sample with [Fe/H] between -0.2 and 0.2 dex \citep{Pinsonneault2018_apokasc2,Elsworth2019_apokasc_evolstate,APOGEE_DR17}. We overlay two [C/N]-mass relations: an empirical relation from \citet{Bufanda2023_CNoutliers} in black, with error bars representing the standard deviation of each bin, and the functional relation from \citet{Roberts2024_CNvsFeH} in blue, with the confidence interval reflecting [Fe/H] between -0.2 and 0.2 dex. We plot the high-mass sample from this work in large circle markers coloured by [Fe/H], using the [C/N] ratios from APOGEE DR17. Note that the uncertainties on [C/N] for the high-mass sample are smaller than the markers.}
    \label{fig:cn_ratio}
\end{figure}

Figure~\ref{fig:cn_ratio} shows the [C/N]-mass relation for our high-mass sample, in comparison to those from \citet{Bufanda2023_CNoutliers} and \citet{Roberts2024_CNvsFeH}, neither of which extend to masses above 3 \Msolar{}.
We compare against stars identified by \citet{Elsworth2019_apokasc_evolstate} as CHeB, using masses from the APOKASC-2 sample \citep{Pinsonneault2018_apokasc2} and [C/N] ratios from APOGEE-DR17 \citep{APOGEE_DR17}. We included APOKASC-2 stars with [Fe/H] between -0.2 and 0.2 dex, matching the metallicity range of our high-mass sample.
In the predictions of models shown in \citet{Roberts2024_CNvsFeH} and in the few observations of high-mass stars, the [C/N]-mass relation becomes flat (i.e. [C/N] is constant) for masses greater than $\sim$2.5 \Msolar{}. Our sample is consistent with this result, although the scatter in [C/N] for our sample is large ($\sim0.3$ dex). The source of this scatter is not immediately clear. It could indicate that more stars in our sample have undergone binary interactions, as predicted by \citet{Bufanda2023_CNoutliers}. However, the scatter of observed [C/N] in ratios in CHeB stars is large for all masses. We have one star (KIC~4372082) with a [C/N] of -0.21, the largest in the sample, which is the most likely to have experienced interactions with a companion according to the results of \citet{Bufanda2023_CNoutliers}. 

Additionally, \citet{Roberts2024_CNvsFeH} predicted a strong relationship between [Fe/H] and [C/N], especially at high-mass. Our data shows an inverse correlation between [C/N] and [Fe/H] (colour coded in Figure~\ref{fig:cn_ratio}), which is even stronger than found previously \citet{Roberts2024_CNvsFeH}. The dependence of [C/N] on [Fe/H] may contribute to the increased scatter of [C/N] in our sample. Ultimately, there is not enough high-mass stars to well constrain the behavior of the [C/N]-[Fe/H]-mass relation above 3 \Msolar{}, but our sample is consistent with previous studies on lower mass stars.


\section{Chromospheric Activity - Ca H \& K}
\label{sec:cahk}



\begin{figure*}
    \centering
    \includegraphics[width=\textwidth]{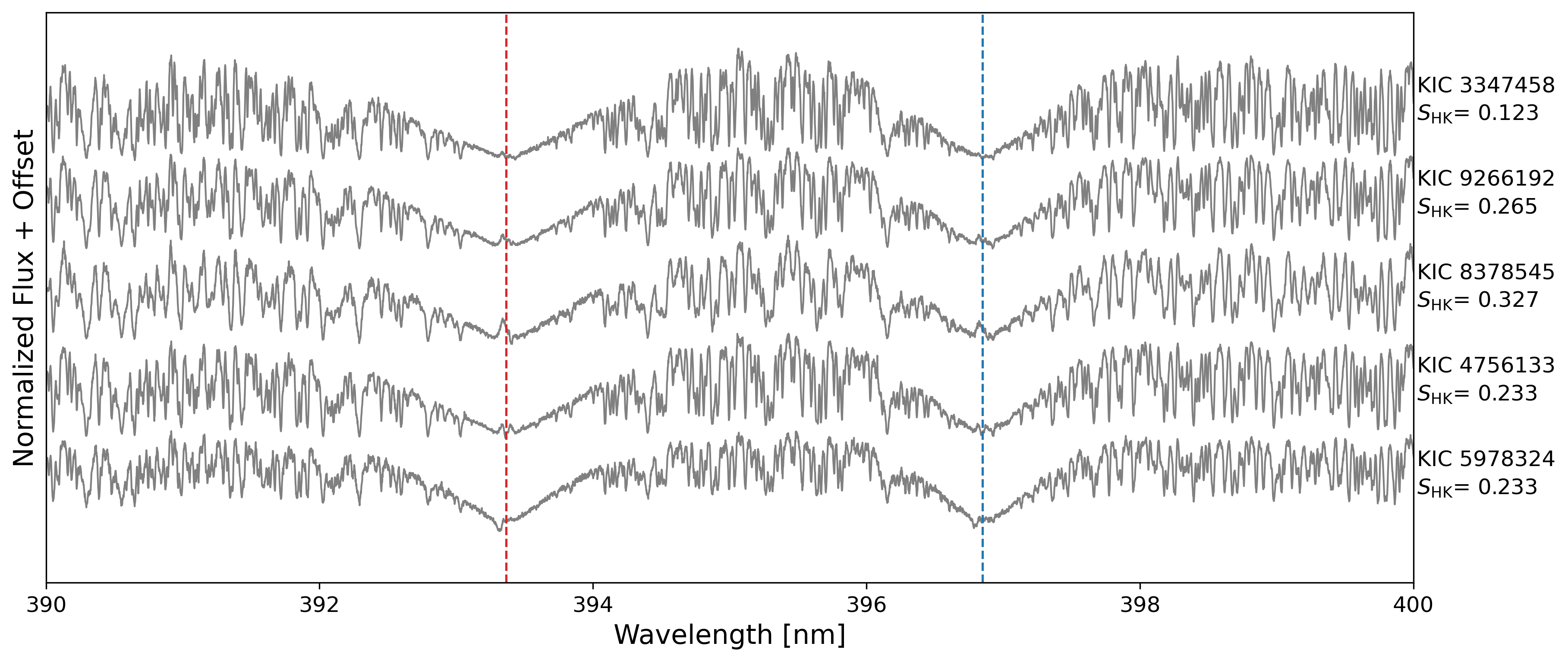}
    \caption{Here we show the Ca H \& K regions of five representative stars from our stellar spectra. The vertical blue dashed line (right) denotes the Ca H line (396.8469 nm) and the vertical red line (left) denotes the Ca K line (393.3663 nm). All spectra are in the rest frame and have been continuum-normalized. On the right we label each star and list the $S$-index for that star.}
    \label{fig:cahk}
\end{figure*}

The Ca II H \& K line cores show emission features in stars with active chromospheres \citep{Wilson1963_cahk_chromosphere,Wilson1968_cahk}. We show the Ca H \& K region of our $b$-band spectra for five representative stars in Figure~\ref{fig:cahk}. A few stars in our sample show the small, double-peaked emission lines indicative of chromospheric activity.
This emission is commonly measured using the $S$-index, defined by the Mt. Wilson H\&K Project \citep[MWO;][]{Duncan1991_MWOobservation} as the flux ratio of the H and K line cores compared to the local continuum:
\begin{equation}
    S = \frac{H + K}{R + V} .
\end{equation}
Here, $H$ and $K$ are the integrated fluxes in a triangular weight centered on the line cores (3968.47 and 3933.66 \AA{}, respectively), with a FWHM of 1.09 \AA{}, while $R$ and $V$ are the fluxes in 20-\AA{} wide regions centered on 3901.068 and 4001.067 \AA{}. Note that since the $S$-index is a flux ratio, it varies from $\sim$0 to $\sim$1, where smaller values indicate lower chromospheric activity levels. This index is sensitive to the response function of the instrument, and therefore we must calibrate our measurements using the Keck/HIRES spectra to those made by the MWO project. We adopted the calibration from \citet{Isaacson2024_s-index}, who used the same spectroscopic setup as our target sample:
\begin{equation}
    S = 22.5 \left( \frac{H + 1.01019 K}{R + 1.26134 V} \right) - 0.006 .
\end{equation}
We note that the above calibration was calculated using main-sequence stars, which have narrower Ca II H \& K chromospheric emission line cores--- commonly referred to as the Wilson-Bappu effect \citep{WilsonBappu1957,Ayres1979_caHKscaling}. This means that some of the chromospheric emission line flux may fall outside of the 1.09 \AA{} width of the integration, systematically lowering the measured $S$-index for all stars in our sample. However, we can safely ignore this effect because it affects all of our stars in the same way, preserving the trends across the sample, and because the overall activity levels are low, and we are thus not missing significant amounts of chromospheric flux outside of the integration window.

We report calibrated stellar $S$-indices for our sample in Table~\ref{tab:massrad_actamp}. The activity levels are low overall, especially compared to highly active close binaries that have an $S$-index $\sim$0.5 or higher from low-resolution spectra \citep{Gehan2022_activity}. 
Although there have been several ensemble studies of red giant stellar activity levels using low-resolution spectra \citep[e.g.][]{Gehan2022_activity,Han2024_lamostrhkgiants}, including those using other types of activity indicators \citep[e.g.][]{Gaulme2020_rotationalmodulation,Han2023_halphaactivity,Gehan2024_chromospheres} and one using the original MWO data \citep{Lehtinen2020_mwogiants}, there are no high-resolution ensemble studies of the $S$-index for red giant stars. Hence, it is difficult to define regimes of strong or weak activity based solely on the $S$-index.
However, in Figure~\ref{fig:cahk} we can see a representative range of the activity levels in the sample. KIC~3347458 shows very weak chromospheric emission, consistent with little or no activity, and has an $S$-index of 0.123. KIC~8378545, on the other hand, has the strongest chromospheric emission in our sample, with an $S$-index of 0.327. 



The $S$-index contains information about the stellar continuum surrounding Ca H \& K. It is therefore sensitive to differences in stellar \Teff{} in main sequence stars of certain \bv{} colours, although this effect is known to be weaker in giant stars (see e.g. \citealt{Mittag2013_rhk_calibrations}). Although the \Teff{} range of our sample is small, there may still be a small effect on the measured $S$-index. We tested this effect by converting the $S$-index to a similar activity index called log(\rhk{}), defined by \citet{Noyes1984_rhkdefinition} as the excess of flux in the Ca II H and K lines due to the chromosphere, normalized by the total bolometric luminosity of the star. In doing so, we found that the overall trend of activity levels for our sample is unchanged. The majority of our stars have weak chromospheric activity regardless of activity indicator, and those stars with large $S$-indices, indicating stronger activity levels, also have log(\rhk{}) consistent with higher activity levels. Further, the usage of $S$-index versus log(\rhk{}) does not change qualitatively the interpretation of the proceeding sections. Additionally, the conversion of $S$-index to log(\rhk{}) requires a considerable number of empirical calibrations that were derived only for main sequence stars, and although there are some calibrations derived for giants \citep{Rutten1984_ccf_bolometric_giants,Mittag2013_rhk_calibrations,GomesdeSilva2021_ccffunction}, these calibrations all rely on \bv{} colour measurements which are not reliably available for our sample. Therefore, we chose to report only the $S$-index for our stars.

\subsection{Activity Cycles and Rotational Modulation}
\label{subsec:activity_cycles}

For our stars, we measured the relative flux in the monthly Kepler Full-Frame Images (FFIs) following the methodology of \citet{Montet2016_ffipipeline} and \citet{Montet2017_ffipipeline} to look for long term brightness variations, which may indicate changes in stellar activity or activity cycles. Alhough we do not see any clear activity cycles, we see 19 stars ($\sim$40\% of the sample) with a detectable change in brightness over the four years of Kepler observations, suggesting a change in their activity over that time period. If we compare to the Sun-like sample of stars studied in \citet{Montet2017_ffipipeline}, where only $\sim$10\% of targets had detectable variability, this suggests that on average these secondary clump stars exhibit larger activity variations on timescales comparable to the four-year duration of the Kepler mission, compared to the longer cycles typical of G dwarfs in a random field sample.


Although we do not see clear indications of full activity cycles, \citet{Gaulme2020_rotationalmodulation} did detect rotational modulation (S$_{\rm ph}$) for 29 of our stars\footnote{Note that \citet{Gaulme2020_rotationalmodulation} only studied Kepler stars that were not dominated by photon noise, so a non-detection in their sample does not necessarily indicate inactivity. Often the missing star is simply faint.} and measured rotation periods for 12 stars. All 12 stars with significant rotational modulation (S$_{\rm ph}$ $>$ 0.05\%) are on the high end of our activity measurements, with $S$-index $>$ 0.15. A comparison of our measured $S$-index with those rotation periods shows a strong correlation, where the most active stars are also the most rapid rotators. This is consistent with the current understanding of magnetic dynamo theory. Such a strong correlation is not seen when using $v\sin i$ instead of rotation period, and we do not observe any correlations of activity or rotation rate with mass. 
For stars with measured rotation periods from \citet{Gaulme2020_rotationalmodulation}, we can add an additional verification of our measurement of $v\sin i$ by comparing it to the rotation period. By using a Kolmogorov-Smirnov (KS) test, we found that the distribution of $\sin i$ in our sample is consistent with a uniform distribution of inclination angles (p-value = 0.38). Given the small sample size, we generated 10,000 simulated samples from a uniform inclination distribution and compared their KS statistics to our data, obtaining an empirical p-value of 0.377. We conclude that our $v\sin i$ measurement is consistent with a random set of inclination angles.

\subsection{Mode Visibility and Activity}
\label{subsec:vis-activity}

We additionally investigated the relationship between the dipole mode visibility ($V^2_{\ell = 1}$, measured in Paper~I) and $S$-index. The former is theorised to be an indicator of the core magnetic field strength \citep{Fuller2015_suppression,Cantiello2016_suppression,Stello2016_visPASA,Stello2016_visNature,Loi2018_magneticsuppression_theory,Muller2025_suppression}, although that explanation is contested \citep{Mosser2017_suppression_mixedmodes}. In the upper panel of Figure~\ref{fig:activity_relations} we show that, for stars with the same \Dnu{}, the dipole mode visibility shows no correlation with $S$-index.
This is not unexpected, even though both are indicators of magnetic fields, because $S$-index probes chromospheric activity, localized to the surface of the star, and (assuming the magnetic greenhouse effect is correct) the dipole mode visibility probes the core magnetic activity. The lack of correlation between the two is consistent with there being little or no connection between the magnetic field near the core of the star and the field near the surface. This is in agreement with the calculations of the ohmic time scale for the core field \citep{Fuller2015_suppression}.

\subsection{Amplitude-Activity Relation}
\label{subsec:amplitude}

Previously, other works have shown that chromospheric activity "suppresses" or decreases the total power of the surface convection-induced acoustic oscillations \citep{Chaplin2011_keplerdetectionsvactivity,Huber2011_testingscalingCHARA,Gaulme2020_rotationalmodulation,Gehan2022_activity,Gehan2024_chromospheres}.
This has been shown directly in the Sun, where the peak heights of its oscillation modes are suppressed during the most active portions of the star's activity cycle due to changes in the surface convective properties, and by extension the damping and excitation rates of the oscillations \citep{Chaplin2000_sunamplitudes_wactivity,Houdek2001_chromosphere_effectsondamping}.
However, the total oscillation power is an ambiguous measurement for our stars because many have suppressed dipole modes, which will decrease the measurement of total oscillation power. Many previous works use the height of a Gaussian fit to the oscillation envelope as a measurement of the total oscillation power, which is subject to effects from lower dipole-mode visibilities \citep{Gaulme2020_rotationalmodulation}. To more directly explore the relationship between surface convection and oscillation amplitudes, we measured the amplitude of only the radial modes, rather than the whole oscillation envelope, because the radial modes are unaffected by the suppression arising from the core.

\begin{figure}
    \centering
    \includegraphics[width=\columnwidth]{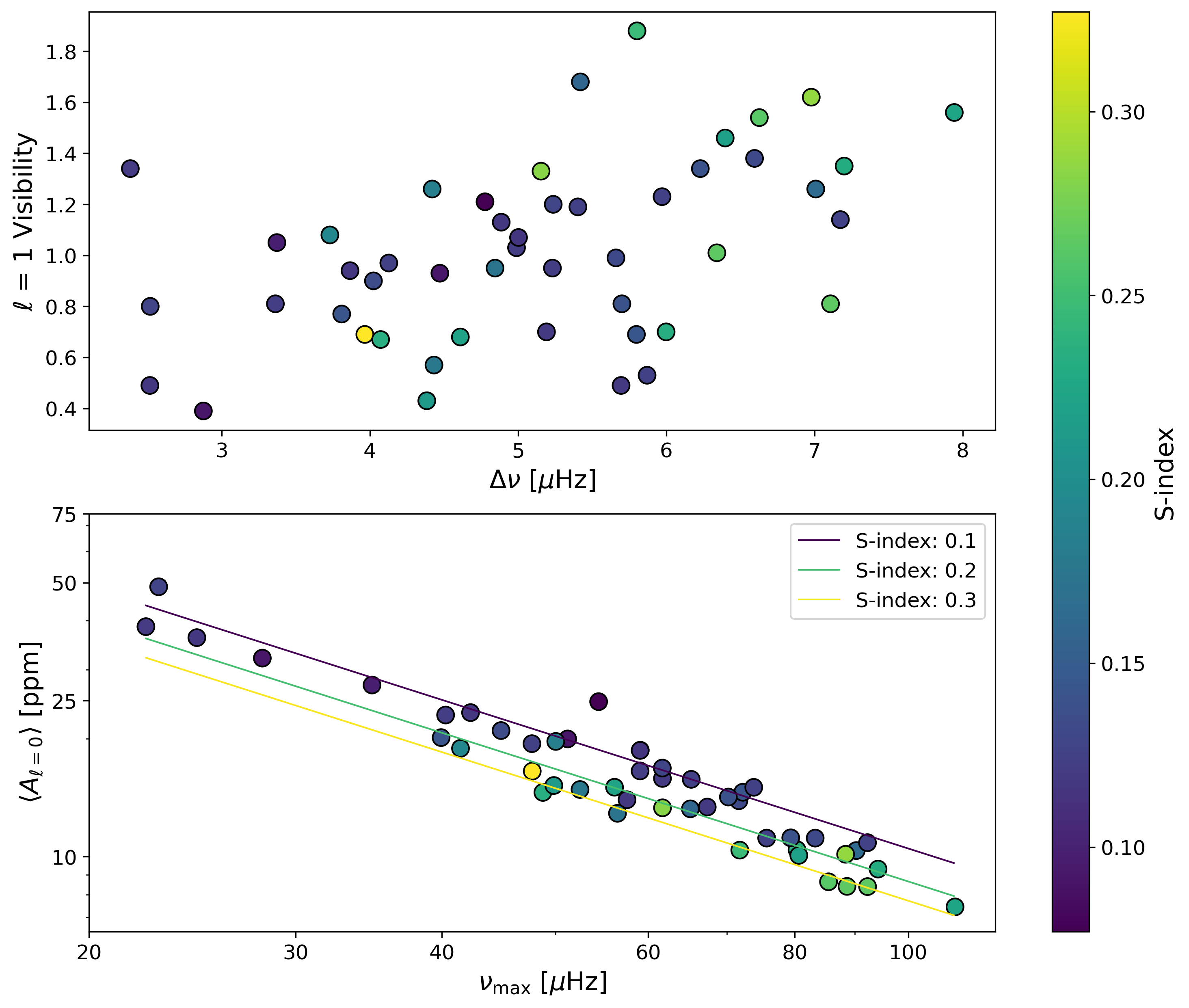}
    \caption{The upper panel shows the dipole mode visibility versus \Dnu{} (both measured in Paper~I), coloured by the $S$-index as indicated by the colour bar on the right. The lower panel shows the average radial mode amplitude (\radamp{}) in ppm versus the \numax{}, coloured by the $S$-index. Additionally, there are three lines, which show the multivariate power law fit to the data (Eqn~\ref{eqn:aa_relation}) for $S$-index values of 0.1 (blue), 0.2 (green), and 0.3 (yellow), which can be interpreted as low, medium, and high levels of activity in the overall sample.}
    \label{fig:activity_relations}
\end{figure}

We calculated the average radial mode amplitude (\radamp{}, in ppm) as 
\begin{equation}\label{eqn:amplitude}
    \langle A_{\ell=0} \rangle = \frac{\sqrt{\left(\sum_{i=1}^{4} P_{\ell=0}\right)/4\Delta\nu}}{\rm sinc \left(\frac{\pi}{2}\frac{\nu_{\rm max}}{\nu_{\rm Nyq}}\right)},
\end{equation}
where $\sum_{i=1}^{4} P_{\ell=0}$ is the integrated power density of the 4 central radial modes using the regions defined by \citet{Stello2016_visPASA}, divided by 4$\Delta\nu$ to convert it to an average radial mode power. We divided the average mode amplitude by a sinc function to correct for the attenuation of signals towards the Nyquist frequency \citep[e.g.,][]{Huber2010_keplerrgs}. We report \radamp{} in Table~\ref{tab:massrad_actamp}.
In the lower panel of Figure~\ref{fig:activity_relations} we show \radamp{} against \numax{} and $S$-index. We see that for stars with the same \numax{}, those with more chromospheric activity also have suppressed radial mode power.
We fitted a power law to this relation of the form
\begin{equation}\label{eqn:aa_relation}
    \langle A_{\ell=0} \rangle = a(\nu_{\rm max})^{-b}(S_{\rm HK})^{c}
\end{equation}
and derived the coefficients a=443.663, b=0.953, and c=0.297. We show the fitted relation for three different values of $S$-index in Figure~\ref{fig:activity_relations} (lower panel). 

As mentioned previously, the effect of surface magnetic activity suppressing oscillations is well known, even with different activity metrics such as the rotational modulation or different line depths \citep{Gaulme2020_rotationalmodulation,Gehan2022_activity,Gehan2024_chromospheres}. However, this is the first time this result has been shown for Kepler stars using explicitly the radial mode amplitudes. In future work, it would be very interesting to expand the measurements of \radamp{} to the larger sample of red giants with measured activity levels. Further exploration of this effect would be useful for efforts on directly simulating the excitation and damping from convection, such as those by \citet{Zhou2020_3Dconvection_excitation_damping}.


\begin{table}
    \centering
    \setlength{\tabcolsep}{5pt}
    \caption{Kinematic Information, including the KIC number, radial velocity (RV) and the Galactic spatial velocities ($U$,  $V$, and $W$). All velocities are expressed in km/s. \label{tab:kinematics}}
    \begin{tabular}{rrrrr}
         KIC & RV (km/s) & $U$ (km/s) & $V$ (km/s) & $W$ (km/s)\\
         \hline
3347458 & -51.34 $\pm$ 0.03 & -26.60 & 204.62 & -14.47 \\
9266192 & -38.85 $\pm$ 0.05 & -21.77 & 216.16 & -9.97 \\
8378545 & -42.35 $\pm$ 0.08 & 235.39 & 153.21 & 20.28 \\
4756133 & -12.80 $\pm$ 0.04 & 76.36 & 215.39 & -8.17 \\
5978324 & 20.84 $\pm$ 0.05 & 71.00 & 252.60 & -4.30 \\
11518639 & -172.04 $\pm$ 0.03 & 318.53 & 44.01 & -50.99 \\
6599955 & -28.35 $\pm$ 0.09 & 82.95 & 202.96 & -30.76 \\
6382830 & -18.19 $\pm$ 0.03 & 17.98 & 226.74 & 5.87 \\
9612933 & -0.11 $\pm$ 0.05 & 65.98 & 241.50 & 0.50 \\
7988900 & 1.68 $\pm$ 0.04 & 65.67 & 240.59 & 1.29 \\
3955502 & -2.59 $\pm$ 0.08 & 138.20 & 199.32 & 12.29 \\
8569885 & -11.91 $\pm$ 0.06 & 16.20 & 235.66 & -8.58 \\
5097690 & -43.25 $\pm$ 0.04 & 1.32 & 208.15 & -23.41 \\
7175316 & -19.02 $\pm$ 0.03 & 28.76 & 218.80 & 0.53 \\
8230626 & -18.74 $\pm$ 0.04 & 42.09 & 222.91 & -11.62 \\
8525150 & -11.09 $\pm$ 0.03 & 24.73 & 235.61 & 10.19 \\
7971558 & -46.87 $\pm$ 0.04 & 88.60 & 180.82 & 0.34 \\
9468199 & -3.37 $\pm$ 0.04 & 65.74 & 232.77 & -0.69 \\
10621713 & -4.04 $\pm$ 0.04 & 150.01 & 223.40 & 16.60 \\
9286851 & -25.56 $\pm$ 0.05 & -6.46 & 226.61 & -3.22 \\
11045134 & -29.03 $\pm$ 0.05 & 30.86 & 217.29 & 2.93 \\
9245283 & -2.60 $\pm$ 0.04 & 63.02 & 239.44 & 0.16 \\
10094550 & -2.20 $\pm$ 0.05 & 42.58 & 244.87 & 18.60 \\
4348593 & 6.71 $\pm$ 0.05 & 36.70 & 243.46 & -0.85 \\
4940439 & -18.58 $\pm$ 0.04 & 20.15 & 227.01 & -5.31 \\
2845610 & -21.81 $\pm$ 0.05 & 41.06 & 209.27 & 2.46 \\
3120567 & 6.60 $\pm$ 0.03 & 56.48 & 242.32 & -21.61 \\
10736390 & -24.34 $\pm$ 0.04 & 29.96 & 226.46 & 1.28 \\
6866251 & 23.03 $\pm$ 0.04 & 40.98 & 254.43 & -0.24 \\
4372082 & -27.07 $\pm$ 0.05 & -18.39 & 231.24 & -12.67 \\
5307930 & -20.63 $\pm$ 0.03 & 66.06 & 212.27 & -11.75 \\
11456735 & -48.41 $\pm$ 0.04 & -22.21 & 195.79 & 30.97 \\
4562675 & 14.74 $\pm$ 0.04 & 21.82 & 257.36 & 3.97 \\
4370592 & 4.79 $\pm$ 0.04 & 40.15 & 244.38 & -6.31 \\
4273491 & 12.66 $\pm$ 0.04 & 81.83 & 236.44 & 10.18 \\
9786910 & -47.70 $\pm$ 0.03 & 12.44 & 198.51 & 6.35 \\
12020628 & -30.66 $\pm$ 0.03 & 2.93 & 219.35 & 2.09 \\
10809272 & -10.81 $\pm$ 0.05 & 157.76 & 212.59 & 18.53 \\
5106376 & -26.39 $\pm$ 0.03 & 47.04 & 206.72 & 11.11 \\
10322513 & -25.35 $\pm$ 0.04 & 21.84 & 230.71 & -38.53 \\
8395466 & -4.24 $\pm$ 0.05 & 54.26 & 237.16 & 0.12 \\
4940935 & -0.75 $\pm$ 0.06 & 51.57 & 237.59 & -10.18 \\
8037930 & -36.76 $\pm$ 0.03 & 100.94 & 193.77 & -20.81 \\
7581399 & -18.87 $\pm$ 0.04 & 68.29 & 218.92 & -37.39 \\
11044315 & -12.02 $\pm$ 0.04 & 55.88 & 232.06 & -2.85 \\
5707338 & -10.86 $\pm$ 0.04 & 86.56 & 208.55 & 16.27 \\
11235672 & -13.39 $\pm$ 0.03 & -2.11 & 234.56 & 7.57 \\
11413158 & 0.82 $\pm$ 0.03 & 51.92 & 245.46 & -9.55 \\
    \end{tabular}
\end{table}

\section{Kinematics and halo orbits}
\label{sec:kinematics}

\begin{figure*}
    \centering
    \includegraphics[width=\textwidth]{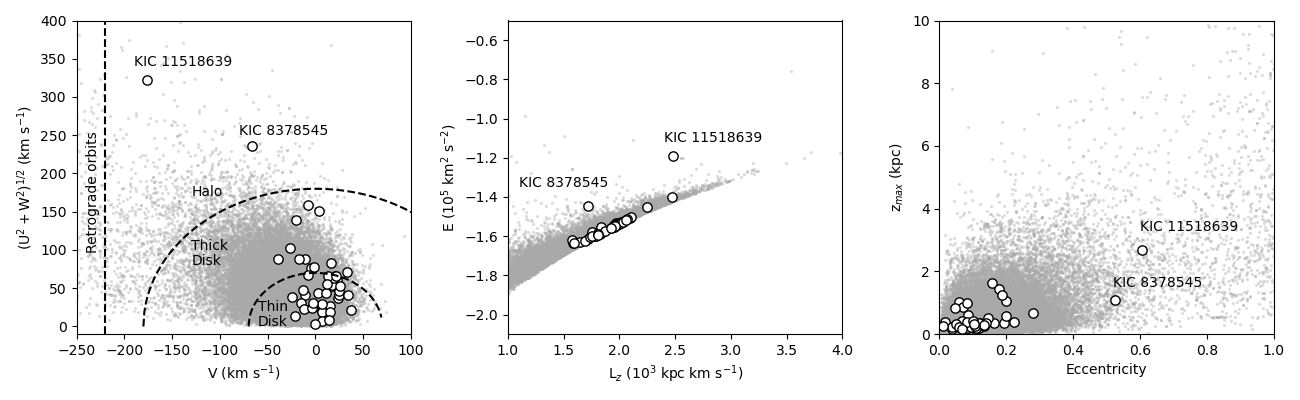}
    \caption{Left panel: Our sample stars (open circles) in the Toomre space of rotational velocity versus perpendicular velocity. Dashed lines show the separations between typical thin disk, thick disk, halo, and retrograde orbits, following the convention from \citet{Bensby14}. Centre panel: Our sample stars in the kinematic space of orbital angular momentum versus total orbital energy, zoomed in on the region of prograde disk rotation. Right panel: Our sample stars in the space of orbital eccentricity versus maximum height above the plane. In all three panels, KIC~11518639 and KIC~8378545 stand out as clearly different from the other stars, and a randomly selected 50~000 stars from the GALAH DR4 data set are shown in the background to illustrate typical Galactic stellar populations.}
    \label{fig:orbits}
\end{figure*}

During our spectroscopic analysis we noticed that one star, KIC~11518639, had a relatively large radial velocity of $-172.04$ km/s\footnote{Note that even though the radial velocity amplitude of the star is large, the shift in its observed frequencies due to the Doppler effect is only 0.032~$\mu$Hz \citep{Davies2014_dopplershift}. The shift therefore has a negligible effect on the measured \numax{} compared to the measurement errors.} as compared to the other stars, which all fall between $-50$ and 50 km/s (see Table~\ref{tab:kinematics}). This star is additionally the most distant star in the sample, at 6.94~$\pm$~0.9~kpc from the Sun. To investigate further, we used the Gaia DR3 astrometry and radial velocity data \citep{GaiaDR3_release} along with photogeometric distances from \citet{BJ21} to calculate velocity components in the Galactocentric frame ($U, V, W$) and orbital parameters ($L_{z}$, $E_{\rm tot}$, eccentriticy, $z_{\rm max}$) for all of our stars. We used \textsc{astropy} \citep{astropy:2013,astropy:2018} to convert sky position, distance, proper motion and radial velocity into Galactocentric velocity components. We then used {\sc galpy} \citep{galpy} with a \citet{McMillan2017_galaxypotential} Galactic potential, a Solar position of (8210, 0, 25) pc in Galactocentric coordinates \citep{McMillan2017_galaxypotential, BHG16}, a circular velocity of 233.1 km s$^{-1}$ and a Solar peculiar velocity of (11.1, 15.17, 7.25) km s$^{-1}$ \citep{SBD10, McMillan2017_galaxypotential} to integrate the orbits forward and calculate the orbital parameters. We list all of the Galactic velocity components and orbital parameters for our stars in Table~\ref{tab:kinematics}. 

Figure~\ref{fig:orbits} shows three views of our stars' orbits, with a random subset of 50~000 stars from the GALAH DR4 data set \citep{Buder25} plotted underneath (small grey circles) to show the typical disk and halo distributions. The left panel shows the Toomre space of rotational versus perpendicular velocity, with the boundaries between the thin/thick disk and the thick disk/halo from \citet{Bensby14} shown as dashed lines. The two most outlying stars, KIC~11518639 and KIC~8378545, are well outside the thick disk region, and several other stars in our sample are well outside the thin disk region, which is quite unexpected for such a young (i.e. high-mass) population \citep[e.g., ][]{Stromberg25}. In Figure~\ref{fig:orbits}, the centre panel shows the vertical component of angular momentum versus the orbital energy, zoomed in on prograde, disk-like rotation. The lower edge of the distribution is bounded by perfectly circular, in-plane orbits, as a circular orbit is the lowest energy orbit at any given angular momentum. The outliers from the Toomre diagram are also well separated in energy from the rest of the group. The right panel shows orbital eccentricity versus z$_{\rm max}$, the maximum height above the disk plane. In this view our two outliers have the largest eccentricities, and the stars in the thick disk region of Toomre space also have larger values of z$_{\rm max}$. 

Galactic star formation typically happens in the disk midplane where the gas density is the highest \citep[e.g., ][]{Walsh2011}. Observational studies have found that the vertical velocity dispersion of a population of disk stars increases with age, which is generally understood to mean that stars form on thin-disk-like orbits and are dynamically heated over time by interactions with spiral arms and giant molecular clouds \citep[e.g., ][]{Hayden17, Sun25}, while the stellar population in the Galactic halo is typically old and metal-poor ($\approx 10$ Gyr, [Fe/H] $\leq -1.0$, e.g. \citealt{Jofre2011_haloage,Mathur2016_keplerhaloRGs}). Our sample of intermediate-mass evolved stars is quite young in Galactic terms: the age of a 2~\Msolar{} red clump star is $\sim$1 Gyr, which serves as a very conservative upper age estimate for our sample \citep{Dotter2016_MIST0,Choi2016_MIST1}. It is surprising, but not unprecedented, to find two of our stars on halo-like orbits and several others on thick disk-like orbits. Studies of Cepheid variables \citep[e.g., ][]{balog97, schmidt04} find that some of the classical Cepheids, which have masses of 3--20 $M_{\odot}$ and correspondingly young ages, can also be found up to 4 kpc from the Galactic plane. There is also a population of young $\alpha$-rich stars that have similar kinematics, although none of the stars in our sample show this characteristic $\alpha$-richness, according to the APOGEE DR17 measurements \citep{Chiappini2015_alpharich,Martig2015_alpharich}.

To understand this finding, we consider two possibilities for the kinematically hotter stars in our sample: that they formed with larger vertical velocities and eccentricities, or that they experienced dynamical interactions within their birth clusters that ejected them into these hotter orbits. Studies of young stellar associations often find that they are dispersing; ``traceback ages'', calculated as the time since the stars were co-located given their current orbits, are one common method for age determination \citep[e.g., ][]{Couture23}. However, the vertical velocity dispersions in young associations are typically at least an order of magnitude smaller than the vertical velocity dispersion in the thick disk, so it is unlikely that the kinematically hotter stars in our sample formed on those orbits.

The second possibility is that both KIC~11518639 and KIC~8378545 were ejected from multiple systems, placing them on their highly peculiar orbits. Three-body interactions are a common source of hypervelocity stars in the Galaxy \citep{li2023_hypervelocitystars} and can also place stars on high-energy bound orbits. To find potential locations where each star was ejected from its original disk orbit, we used {\sc galpy}, with the same starting conditions as before, to integrate each star's orbit backward in time in 10-Myr steps to locate its most recent crossings of the Galactic midplane. For KIC~11518639 we found six disk crossings in the past Gyr, with the most recent occurring 240 Myr ago at a Galactocentric radius of 18.45 kpc. This location seems much too far out in the disk to be a site for recent star formation, so we continued to follow the orbit and found the next previous midplane crossing at 320 Myr ago and 7.4 kpc from the Galactic centre, which seems a plausible ejection location. The midplane crossing before that was 600 Myr ago at 13.63 kpc, again an unlikely location for recent star formation. For KIC~8378545 we found 13 midplane crossings in the past Gyr, with several occurring at a Galactocentric distance between 4 and 6 kpc. Any of these could have been the launching point for the current orbit of KIC~8378545. However, it is worth mentioning that integrating orbits is a complicated calculation, with the errors compounding with increasing trace-back time. While the ejection scenario seems most likely explanation for the kinematics of these two stars, more careful modelling of the potential in the disk would be needed to track their orbits precisely.

\section{Summary}
\label{sec:conclusions}

In this work, we have acquired Keck HIRES spectra for our sample of 48 Kepler high-mass red giants to explore various properties. Our foremost goal with these spectra was to homogenize the \Teff{} estimates for each star, such that our future work can test the asteroseismic scaling relations. We used the python tool \texttt{iSpec} to perform synthetic spectrum synthesis using the SPECTRUM code \citep{GrayCorbally1994_SPECTRUM} in conjunction with MARCS model atmospheres \citep{Gustafsson2008_MARCSmodels} to determine the stellar parameters for each star, namely \Teff{}, \logg{}, [Fe/H], and $v\sin i$. These values are consistent with the values measured in the literature.


Since we intend to use this sample to test the asteroseismic mass estimates from scaling relations, it is paramount that we understand which stars are likely to have been influenced by binary evolution and which stars are single stars that will be most similar to stellar model predictions.
Therefore, we investigated two potential binarity indicators: the presence of a strong Li 6707 \AA{} line and an anomalous [C/N] value. The former is theorised to arise from binary tidal spin-up interactions which warm the burning region and create excess Li \citep{Casey2019_lirichgiants}, and outliers in the latter have been shown to be more common in binary systems \citep{Bufanda2023_CNoutliers}. In our sample, we do not see any stars with Li-enrichment, but we do find one (KIC~ 437208) with a large [C/N] value of $-0.21$.
We do not see any clear evidence of binary interactions using these measurements.
See Paper~I for additional exploration of binarity indicators, including ``anomalous peaks'' \citep{Colman2017_anamolouspeaks}, Kepler pixel data inspection, Kepler crowding metrics, the Gaia RUWE parameter, and crossmatches with non-single star catalogues.

We used the HIRES $b$-band spectra to measure the chromospheric activity indicators $S$-index and $\log(R'_{\rm HK}$). 
A few stars show evidence of chromospheric activity, but the majority of our sample are not especially active and have $S$-index values consistent with typical activity levels of red giant stars \citep[e.g.][]{Gehan2022_activity}.
We did not find clear evidence of activity cycles in the Kepler data.
We compared the activity levels to two asteroseismic parameters that are often employed to describe stellar magnetic fields. First, we compared to the visibility of the dipole modes ($V^2_{\ell = 1}$, measured in Paper~I), which may correlate with the presence of interior magnetic fields \citep{Fuller2015_suppression,Stello2016_visPASA,Stello2016_visNature,Cantiello2016_suppression,Mosser2017_suppression_mixedmodes,Loi2018_magneticsuppression_theory,Muller2025_suppression}. We saw no correlation between stellar chromospheric activity and the dipole mode visibilities, which implies that there is no relationship between the magnetic fields near the stellar surface and those near the core. We additionally confirmed that increased chromospheric activity suppresses the strength of the oscillations.
This was known previously \citep{Gaulme2020_rotationalmodulation,Gehan2022_activity,Gehan2024_chromospheres} but this is the first time that the radial modes have been isolated, rather than using the total oscillation power. Since the radial modes are not suppressed by the core magnetic fields, it is an unambiguous detection of the effects of the chromosphere in mode excitation and damping \citep{Houdek2001_chromosphere_effectsondamping,Li2021_magfields_surfaceeffect}.

Finally, we used astrometry and radial velocities from Gaia DR3 \citep{GaiaDR3_release} to explore the kinematics of our sample. The majority of our stars are following thin disk orbits, but some have thick disk kinematics and two (KIC~11518639 and KIC~8378545) are clearly on halo-like orbits. 
This is quite unexpected, since the estimated age of the Galactic halo \citep[$\sim$10 Gyr, ][]{Jofre2011_haloage,Kilic2019_haloage} is significantly older than even the most conservative age estimates for these stars \citep[$\sim$1 Gyr, ][]{Dotter2016_MIST0,Choi2016_MIST1}. 
We believe that the most likely explanation is dynamical ejection from a multiple system. We used {\sc galpy} to integrate the two halo star orbits backward in time to find plausible midplane crossings at which ejection could have happened. The kinematically warm orbits of our sample raise questions about the accuracy and viability of kinematics in unravelling Galactic history.

The high-mass Kepler red giants studied here have proven to be remarkably typical stars. We do not find any especially unusual spectroscopic parameters, obvious binarity, or activity levels. Notwithstanding their unusual Galactic orbits, these stars fit well with the our knowledge of the overall sample of Kepler red giants. In our next work, we will perform individual frequency modelling on these stars in conjunction with the new homogeneously measured spectroscopic parameters found in this work. This boutique modelling scheme will yield independent estimates of stellar masses and radii, enabling us to evaluate the adherence to asteroseismic scaling relations in the rare, often-excluded high-mass red giants. In addition, boutique modelling may allow us to measure the inclinations and, potentially, core/envelope rotation rates of stars in a mass range not previously covered \citep{LiGang2024_core_env_rotation}, which may help to constrain angular momentum evolutionary models \citep[e.g.][]{Fuller2019_taylerspruit,Eggenberger2022_angmomtransport}.


\section*{Acknowledgements}

Thank you to Sven Buder and Alexander Ji for their helpful advice on the spectroscopy involved in this paper. We additionally thank the anonymous reviewer for their time and effort in improving this article. CC and TB gratefully acknowledge support from the Australian Research Council through Discovery Project DP210103119 and Laureate Fellowship FL220100117. DS acknowledges support from ARC DP190100666. SLM acknowledges support from ARC DP220102254 and the UNSW Scientia Fellowship programme. L.M.W. acknowledges support from the NASA Exoplanet Research Program (grant no. 80NSSC23K0269). 

Some of the data presented herein were obtained at Keck Observatory, which is a private 501(c)3 non-profit organization operated as a scientific partnership among the California Institute of Technology, the University of California, and the National Aeronautics and Space Administration. The Observatory was made possible by the generous financial support of the W. M. Keck Foundation.
The authors wish to recognize and acknowledge the very significant cultural role and reverence that the summit of Mauna Kea has always had within the Native Hawaiian community. We are most fortunate to have the opportunity to conduct observations from this mountain.

This work made use of several publicly available {\tt python} packages: {\tt astropy} \citep{astropy:2013,astropy:2018}, 
{\tt lightkurve} \citep{lightkurve2018},
{\tt matplotlib} \citep{matplotlib2007}, 
{\tt numpy} \citep{numpy2020},
{\tt pandas} \citep{mckinney-proc-scipy-2010_pandas,the_pandas_development_team_2024_13819579}, 
{\tt scipy} \citep{scipy2020}, {\tt astroquery} \citep{astroquery} and {\tt galpy} \citep{galpy}.


\section*{Data Availability}

All tables present in this article can be downloaded freely via Zenodo. The spectra measured for this work may be requested from the corresponding author.


\typeout{}
\bibliographystyle{mnras}
\bibliography{highmass_rc}

\bsp	
\label{lastpage}
\end{document}

%% file: definitions.tex
\newcommand{\mmode}[1]{\ifmmode{#1}\else{$#1$}\fi}

\newcommand{\Teff}[0]{\mmode{T_\text{eff}}}

\newcommand{\Msolar}[0]{\mmode{\text{M}_{\odot}}}

\newcommand{\logg}[0]{\mmode{\log g}}
\newcommand{\Dnu}[0]{\mmode{\Delta\nu}}

\newcommand{\numax}[0]{\mmode{\nu_\text{max}}}

\usepackage{CJKutf8}
\newcommand{\CNnames}[1]{{\begin{CJK}{UTF8}{gbsn}~(#1)~\end{CJK}}}

